\documentclass[journal]{IEEEtran}

\usepackage{color}
\usepackage{amsmath}
\usepackage{amssymb}
\usepackage{graphicx}
\usepackage{float}
\usepackage{fixltx2e}
\usepackage{subeqnarray}
\usepackage{cases}
\allowdisplaybreaks
\IEEEoverridecommandlockouts
\usepackage[square, comma, sort&compress, numbers]{natbib}
\usepackage{multirow}
\usepackage{stfloats}
\usepackage{fixltx2e}
\usepackage[font={small}]{caption}
\usepackage{epsfig}
\usepackage{epstopdf}
\usepackage{optidef}

\usepackage{algorithm}
\usepackage{algpseudocode}
\ifCLASSINFOpdf
\else
\fi
\usepackage{amsmath}
\usepackage{array}
\usepackage{fixltx2e}

\usepackage{stfloats}

\begin{document}

\title{Trellis-Coded Non-Orthogonal Multiple Access\\}

\author{\IEEEauthorblockN{Xun Zou, \textit{Student Member, IEEE}, Mehdi~Ganji, \textit{Student Member, IEEE}, and Hamid~Jafarkhani, \textit{Fellow, IEEE}}\\
	\thanks{This work was supported in part by the NSF Award CCF-1526780. The authors are with the Center for Pervasive Communications and Computing, Department of Electrical Engineering and Computer Science, University of California, Irvine, CA, 92697 USA (email: \{xzou4, mganji, hamidj\}@uci.edu).}
}

\maketitle
\begin{abstract}
In this letter, we propose a trellis-coded non-orthogonal multiple access (NOMA) scheme. The signals for different users are produced by trellis coded modulation (TCM) and then superimposed on different power levels. By interpreting the encoding process via the tensor product of trellises, we introduce a joint detection method based on the Viterbi algorithm. Then, we determine the optimal power allocation between the two users by maximizing the free distance of the tensor product trellis. Finally, we manifest that the trellis-coded NOMA outperforms the uncoded NOMA at high signal-to-noise ratio (SNR).
\end{abstract}

\begin{IEEEkeywords}
Non-orthogonal multiple access, trellis coded modulation, superposition coding.
\end{IEEEkeywords}

\section{Introduction}
Non-orthogonal multiple access (NOMA) is envisaged as one of the potential technologies in the next generation wireless communication systems. Users in NOMA systems can share the non-orthogonal resources, e.g., the frequency spectrum and the time slot. From a unified perspective, NOMA consists of code-domain NOMA and power-domain NOMA~\cite{wang2018non}. 

Both code-domain and power-domain NOMA have been extensively studied in the existing literature. In the power-domain NOMA systems, the signals of different users are assigned different powers. Then, one major challenge is the optimal power allocation as discussed, for example, in~\cite{liu2017downlink,choi2016power}. The optimal power can be determined according to the channel conditions to maximize users' achievable rates. Superposition coding and successive interference cancellation (SIC) techniques are utilized at the transmitter and the receiver, respectively. Again, there are many studies on how to perform these techniques efficiently, for example~\cite{vanka2012superposition,zhang2011unified}. 

The code-domain NOMA has its origin in code division multiple access (CDMA), including sparse code multiple access (SCMA)~\cite{nikopour2013sparse} and trellis coded multiple access (TCMA)~\cite{aulin1999trellis}. The signals of multiple users are separated by user-specific features, e.g., the uniquely assigned codeword of each user. In the code-domain NOMA, the main efforts are devoted to the multi-user detection, for example, the design of multidimensional constellations~\cite{nikopour2013sparse,di2019tcm}. To the best of our knowledge, the joint design of the code-domain and power-domain NOMA has never been studied.

In this work, we apply trellis coded modulation (TCM) to the power-domain NOMA, taking advantages of the coding gain and the power optimization. Utilizing superposition coding, the signals for multiple users are superimposed on different power levels. Compared with~\cite{di2019tcm}, the main contribution of this work is introducing the power allocation to code-domain NOMA. The performance can be improved by allocating proper powers to the signals of different users. Instead of utilizing TCM purely for codeword design in~\cite{di2019tcm}, TCM is employed in this work to jointly optimize the error control coding and modulation. Therefore, the Viterbi algorithm can be directly applied to the proposed scheme. By interpreting the modulating process via the tensor product of trellises~\cite{jafarkhani1999design,jafarkhani1999multiple}, we implement the maximum likelihood sequence detection (MLSD) based on the Viterbi algorithm~\cite{viterbi1971convolutional}. Furthermore, we derive the optimal power allocation between the two users by maximizing the free distance of the tensor product trellis.

The key difference between the trellis-coded NOMA and the traditional TCMA lies in the multiple access scheme. In TCMA, the signals of multiple users are differentiated by their unique features, for example, convolutional encoder, constellation, or interleaver~\cite{brannstrom2002iterative}. However, in the trellis-coded NOMA, the signals are differentiated only by the power levels. Furthermore, for the first time, we provide insight into the power optimization for the superimposed TCM signals.

\section{System Model}\label{sec_sys_model}
\begin{figure}[t b]
	\centering
	\includegraphics[width=3in]{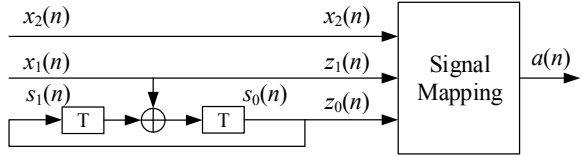}
	\caption{Illustration of an 8-PSK 4-state TCM encoder.}
	\label{fig_conv_encoder}
\end{figure}

\begin{figure}[t b]
	\centering
	\includegraphics[width=3.5in]{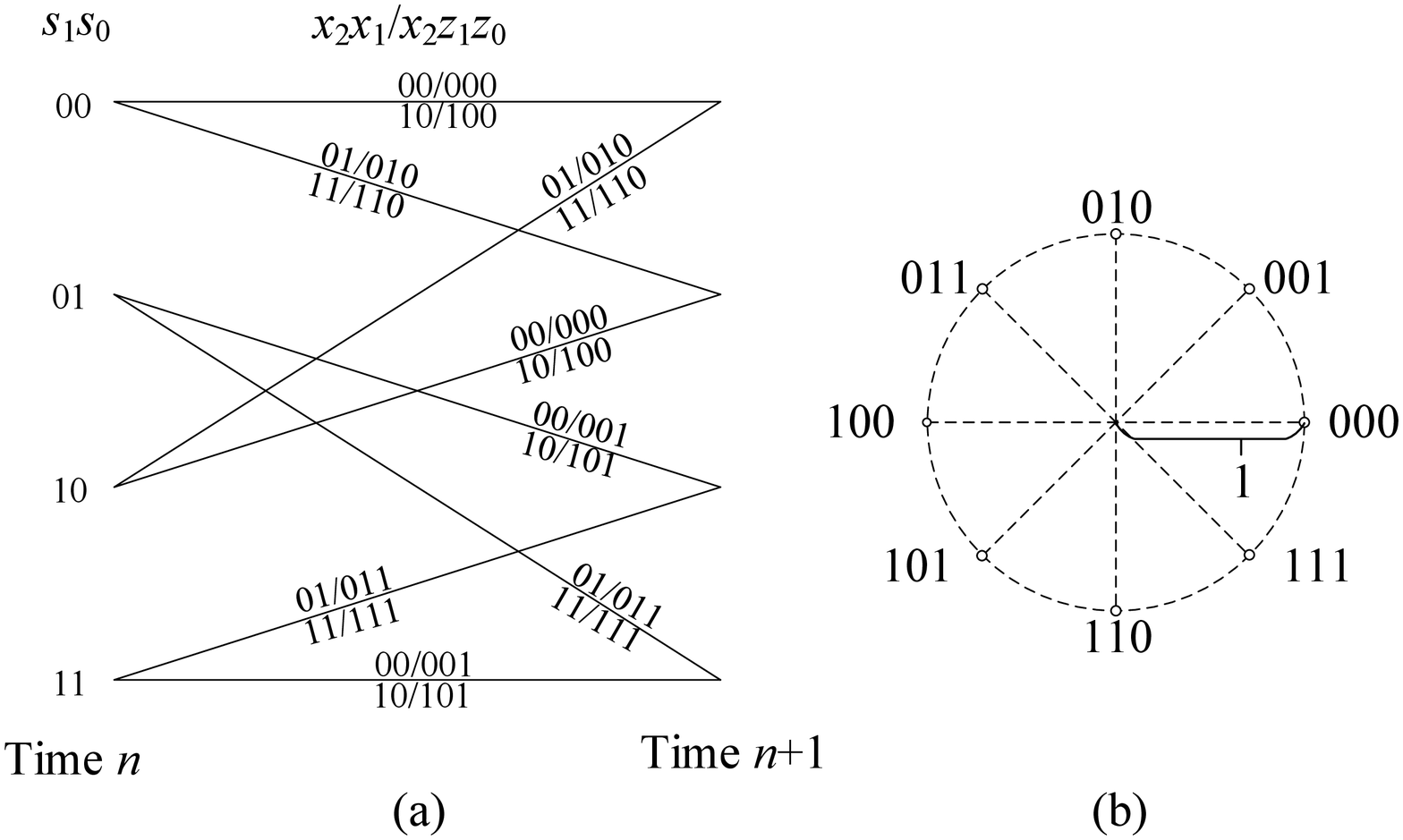}
	\caption{(a) Trellis representation of 8-PSK 4-state TCM. (b) The mapping of 8-PSK constellation.}
	\label{fig_4_state_trellis}
\end{figure}

In this letter, we consider a downlink NOMA system consisting of one base station (BS) and two users. Superposition coding is employed at the transmitter. The power allocated to User~$i$'s signal is denoted as $P_i$, $i = 1, 2$. The channel coefficient between the BS and User~$i$ is represented by $h_i$. We adopt the block fading channel model, i.e., the channel remains static within each block and changes independently from one block to another~\cite{liu2017downlink,choi2016power}. We assume that the channel state information is perfectly known by the BS and users. Without loss of generality, we assume that $|h_1|^2 > |h_2|^2$. To stipulate the user fairness, we set $P_2 > P_1$. 
In what follows, the 8-phase-shift keying (PSK) 4-state TCM serves as an example of TCM~\cite{ungerboeck1987trellis}, which is depicted in Fig.~\ref{fig_conv_encoder}. The trellis diagram and the 8-PSK mapping are shown in Figs.~\ref{fig_4_state_trellis} (a) and (b), respectively. In Figs.~\ref{fig_conv_encoder} and \ref{fig_4_state_trellis}, $x_1$ and $x_2$ represent the uncoded bits while $z_0$ and $z_1$ denote the coded bits via the convolutional encoder. For the sake of brevity, we employ the signal constellation with unit signal power, i.e., $\mathrm{E_b} = 1$. Note that the proposed scheme can be applied to the case where two users employ different modulations/trellises and also the case of more than two users.

In the proposed trellis-coded NOMA, the signals for Users~1 and 2 are first modulated by TCM, as shown in Figs.~\ref{fig_conv_encoder} and \ref{fig_4_state_trellis}, and then superimposed on different power levels. Using superposition coding, the $n$th transmitted symbol at the BS is given by $\sqrt{P_1} a_1(n) + \sqrt{P_2} a_2(n)$ where $a_i(n)$ is the $n$th symbol for User~$i$ after TCM. Then, the $n$th received sample at User~$i$ is given by
\begin{align}\label{eq_superposition_coding}
    y_i(n) = h_i\left[\sqrt{P_1} a_1(n) + \sqrt{P_2} a_2(n)\right] + w_i(n),
\end{align}
where $w_i(n)\sim \mathcal{CN}(0, \sigma_i^2)$ is the additive noise. At users, the modulated symbols are detected and then the binary information bits are recovered from the modulated symbols, which will be explained in the next section.


\section{Tensor Product of Trellises and Detection Design}\label{sec_tensor_product_and_detector}
In this section, we first present the separate detection method with SIC. Then, we propose the joint detection method based on a novel trellis structure known as ``tensor product of trellises''.

\subsection{Separate Detection with SIC}
In the separate detection scheme, the signals for Users~1 and 2 are detected separately. User~2 (the weak user) detects its own signal by considering User~1's signal as noise. User~1 (the strong user) utilizes SIC, i.e., first detects User~2's signal, removes it from the superimposed signal, and then detects its own signal. The Viterbi algorithm~\cite{viterbi1971convolutional} can be employed to determine the sequence with the minimum Euclidean distance from the received sequence using the 4-state trellis in Fig.~\ref{fig_4_state_trellis}~(a).


\begin{figure}[t b]
	\centering
	\includegraphics[width=2.5in]{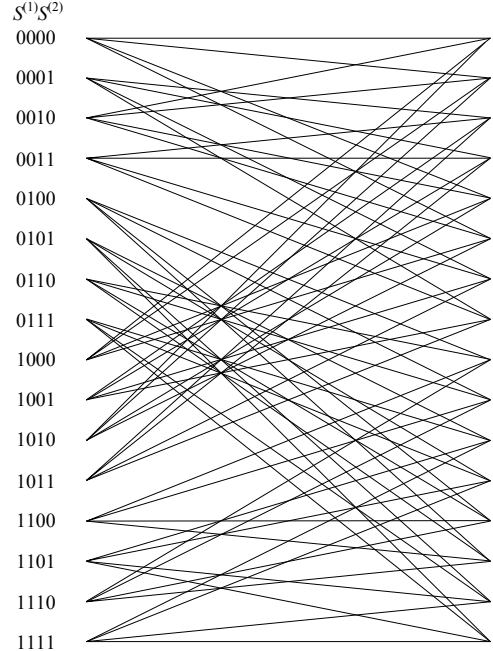}
	\caption{Underlying tensor product of trellises.}
	\label{fig_16_state_trellis}
\end{figure}

\subsection{Joint Detection with Tensor Product of Trellises}\label{subsec_tensor_product}
First, we review the concept of the tensor product of trellises~\cite{jafarkhani1999multiple,jafarkhani1999design}. Let us consider trellises $T_1$ and $T_2$ with $r_1$ and $r_2$ states, respectively, and $S_i^{(l)}$, $i=1,\cdots,r_l$, denotes the $i$th state of $T_l$. The tensor product of $T_1$ and $T_2$, denoted as $T_1 \otimes T_2$, can be represented as a trellis with $r_1\times r_2$ states. Each state in $T_1 \otimes T_2$ is given by $S_i^{(1)} S_j^{(2)}$, $i = 1, \cdots, r_1$, $j = 1, \cdots, r_2$. The state transition from $S_i^{(1)} S_j^{(2)}$ to $S_k^{(1)} S_l^{(2)}$ exists if and only if there exist transitions from $S_i^{(1)}$ to $S_k^{(1)}$ in $T_1$ and from $S_j^{(2)}$ to $S_l^{(2)}$ in $T_2$. One can easily extend the definition of the tensor product trellis to the case of more than two trellises.

Let us revisit the modulating process of two users' signals in Section~\ref{sec_sys_model}. The symbols for Users~1 and 2 are modulated independently through the 4-state trellis, shown in Fig.~\ref{fig_4_state_trellis} (a). Let $T_1$ and $T_2$ stand for the trellises employed to modulate the symbols for Users~1 and 2, respectively. The tensor product trellis $T_1 \otimes T_2$ is the 16-state trellis in Fig.~\ref{fig_16_state_trellis}. Every pair of state transitions in $T_1$ and $T_2$ can be represented by a unique transition path in $T_1 \otimes T_2$. For example, let us assume that the state of $T_1$ transits from $S_i^{(1)}$ to $S_k^{(1)}$ producing the modulated symbol $a_1$ and the state of $T_2$ transits from $S_j^{(2)}$ to $S_l^{(2)}$ generating the modulated symbol $a_2$. From the perspective of $T_1 \otimes T_2$, the state transits from $S_i^{(1)} S_j^{(2)}$ to $S_k^{(1)} S_l^{(2)}$ and the superimposed symbol $\sqrt{P_1}a_1 + \sqrt{P_2}a_2$ is produced. Since every state transition can be realized by two parallel paths in $T_1$ and $T_2$, as shown in Fig.~\ref{fig_4_state_trellis} (a), every state transition in $T_1 \otimes T_2$ includes $2\times 2 = 4$ parallel paths.

The description of the tensor product trellis demonstrates the equivalence of the trellis-coded NOMA and the TCM using the tensor product trellis. The joint detection is to detect both users' signals jointly by treating the trellis-coded NOMA as a regular TCM with the tensor product trellis. In the joint detection, the Viterbi algorithm is implemented using the tensor product trellis. It is worth mentioning that there is no necessity to modulate the signals for Users~1 and 2 jointly using the tensor product trellis at the transmitter. The transmitted symbols for each user can be modulated independently according to its own trellis by applying an appropriate power allocation scheme to ensure a good decoding performance (as shown in Section~V). 

Since the Viterbi algorithm can be employed in joint decoding, the computational complexity increases linearly with the number of decoded symbols, $N$. More specifically, if the number of states in $T_i$ ($i=1, 2$) is $K_i$ and the total number of edges in $T_i$ is $L_i$, the computational complexity of the joint detection method is given by $O(N(K_1K_2 + L_1L_2))$ while that of the separate detection method with SIC is $O(N(K_1 + K_2 + L_1 + L_2))$.

\section{Power Optimization}\label{sec_power_ratio}
In this section, we study the power allocation to optimize the performance of the joint detection scheme. The power allocation is optimized under two power constraints. One is the sum power constraint, i.e., $P_1 + P_2 \le P$ where $P$ is the total transmit power. The other constraint is $P_1 < P_2$ which is added with no loss of generality. We adopt the free distance of the tensor product trellis, $d_{\mathrm{free}}$, to measure the performance, which is widely used in the existing TCM studies, for example~\cite{benedetto1999principles}. A larger free distance results in a better performance at high signal-to-noise ratio (SNR). As will be illustrated later, the free distance is a function of the power coefficients $P_1$ and $P_2$. We obtain the optimal powers by maximizing the free distance.

The free distance is defined as the minimum Euclidean distance between any pair of valid and distinct sequences produced by a given trellis, i.e.,
$d_{\mathrm{free}} = \mathop{\arg\min}_{\mathbf{a}_1, \mathbf{a}_2 \in V, \mathbf{a}_1\neq\mathbf{a}_2} ||\mathbf{a}_1 - \mathbf{a}_2||$ where $V$ is the set of all valid sequences. The free distance can be determined by choosing the minimum of two candidates: the minimum Euclidean distance between the symbols produced by the parallel paths, i.e., $d_{\mathrm{parallel}}$, and that between the sequences which diverge from the same state and then merge at the same state, i.e., $d_{\mathrm{D\&M}}$. The subscript $\mathrm{D\&M}$ is the acronym for ``diverging and merging''. In what follows, we analyze these two distances separately. Assume that there are two different paths in $T_1 \otimes T_2$ producing $\sqrt{P_1}u_1 + \sqrt{P_2}v_1$ and $\sqrt{P_1}u_2 + \sqrt{P_2}v_2$, where $u_1$ and $u_2$ are the modulated symbols of $T_1$ and $v_1$ and $v_2$ are those of $T_2$.
\vspace{-2mm}
\subsection{Parallel Paths}\label{subsubsec_paralel_path}
\begin{figure}[t b]
	\centering
	\includegraphics[width=2.8in]{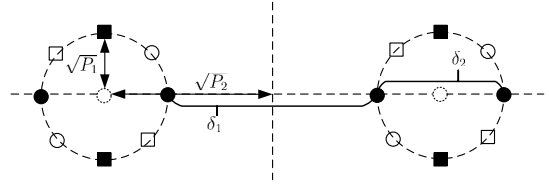}
	\caption{Illustration of the minimum Euclidean distance in the superimposed constellation.}
	\label{fig_parallel_dist}
\end{figure}

First, we study the case where $\sqrt{P_1}u_1 + \sqrt{P_2}v_1$ and $\sqrt{P_1}u_2 + \sqrt{P_2}v_2$ are produced by the parallel paths in $T_1\otimes T_2$. 
Fig.~\ref{fig_parallel_dist} illustrates the possible positions of $\sqrt{P_1}u_1 + \sqrt{P_2}v_1$ and $\sqrt{P_1}u_2 + \sqrt{P_2}v_2$ in the superimposed constellation when $v_1$ and $v_2$ are chosen from $\{1, -1\}$. Because of symmetry, the minimum Euclidean distance for all the other choices will be the same. In Fig.~\ref{fig_parallel_dist}, there are four different markers, hollow/solid square/circle. The superimposed symbols depicted by the same marker are the symbols produced by the parallel paths for a specific state transition in $T_1\otimes T_2$. Every state transition in $T_1\otimes T_2$ can be realized by four parallel paths. Therefore, there are four positions for every marker. The minimum Euclidean distance between parallel paths can be found by calculating the Euclidean distance between the points sharing the same marker. It is clear from Fig.~\ref{fig_parallel_dist} that the minimum Euclidean distance is either $\delta_1$ or $\delta_2$. Thus,
\begin{align}\label{eq_d_parallel}
    d_{\mathrm{parallel}} \!=\! \min\left\{\delta_1, \delta_2\right\} \!=\! \min\left\{2\sqrt{P_2} \!-\! 2\sqrt{P_1}, 2\sqrt{P_1}\right\}.
\end{align}

\begin{figure}[t b]
	\centering
	\includegraphics[width=2.8in]{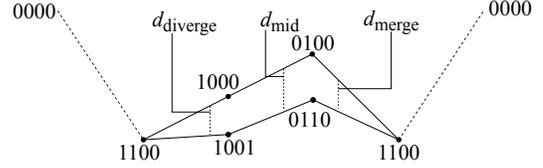}
	\caption{Illustration of the diverging-and-merging paths with the minimum Euclidean distance.}
	\label{fig_diverge_merge_paths}
\end{figure}
\vspace{-5mm}
\subsection{Diverging-and-Merging Paths}
Second, we study the Euclidean distance between the sequences which diverge from the same state and then merge at the same state. It can be shown that if two sequences diverge from any state, it takes at least three transitions to merge at the same state. We utilize the exhaustive search to find a pair of sequences with the minimum Euclidean distance among all pairs of distinct sequences, which is shown in Fig.~\ref{fig_diverge_merge_paths}. Note that all valid codewords start and end at state zero. However, any common sub-sequence will not contribute to $d_\mathrm{free}$. Therefore, to calculate $d_\mathrm{free}$ in Fig.~\ref{fig_diverge_merge_paths}, we need to consider the state transitions $1100\rightarrow 1000 \rightarrow 0100 \rightarrow 1100$ and $1100\rightarrow 1001 \rightarrow 0110 \rightarrow 1100$. As shown in Fig.~\ref{fig_diverge_merge_paths}, the squared Euclidean distance between the diverging-and-merging paths is given by
\begin{align}
    d_{\mathrm{D\&M}}^2 = d_{\mathrm{diverge}}^2 + d_{\mathrm{mid}}^2 + d_{\mathrm{merge}}^2.
\end{align}

\begin{figure}[t b]
	\centering
	\includegraphics[width=2.2in]{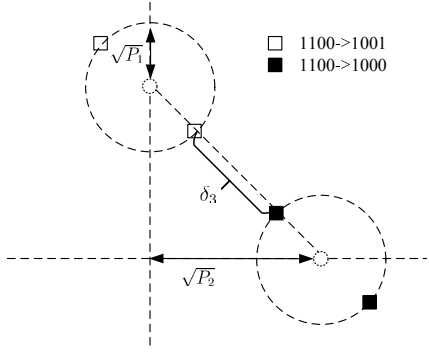}
	\caption{Illustration of the minimum Euclidean distance between the symbols produced by diverging paths.}
	\label{fig_diverge_merge_branch}
\end{figure}

First, let us focus on the diverging paths in Fig.~\ref{fig_diverge_merge_paths}. According to Fig.~\ref{fig_4_state_trellis}, the superimposed symbol produced by the path $1100\rightarrow 1000$ is given by $\sqrt{P_1}u_1 + \sqrt{P_2}v_1$, where $u_1\in \{e^{j3\pi/4}, e^{j7\pi/4}\}$ and $v_1\in \{1, -1\}$. Similarly, the superimposed symbol produced by $1100\rightarrow 1001$ is given by $\sqrt{P_1}u_2 + \sqrt{P_2}v_2$, where $u_2\in \{e^{j3\pi/4}, e^{j7\pi/4}\}$ and $v_2\in \{e^{j\pi/2}, e^{j3\pi/2}\}$. The positions of the superimposed symbols can be shown in Fig.~\ref{fig_diverge_merge_branch}. The minimum Euclidean distance between the diverging paths is given by
\begin{align*}
    d_{\mathrm{diverge}} = \delta_3 = |\sqrt{2P_2} - 2\sqrt{P_1}|.
\end{align*}
One can employ the same approach to derive the minimum Euclidean distance between the merging paths and find that $d_{\mathrm{merge}} = d_{\mathrm{diverge}}$.

\begin{figure}[t b]
	\centering
	\includegraphics[width=2.2in]{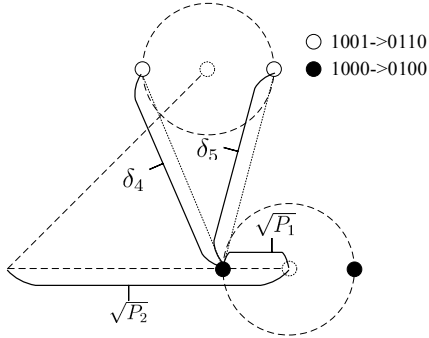}
	\caption{Illustration of the minimum Euclidean distance between the symbols in the intermediate stage of the diverging-and-merging paths.}
	\label{fig_mid_branch}
\end{figure}

Second, we investigate the Euclidean distance $d_{\mathrm{mid}}$ in  Fig.~\ref{fig_diverge_merge_paths}. The superimposed symbol produced by the path $1000\rightarrow 0100$ is given by $\sqrt{P_1}u_1 + \sqrt{P_2}v_1$, where $u_1\in \{1, -1\}$ and $v_1\in \{1, -1\}$. Similarly, the superimposed symbol produced by the path $1001\rightarrow 0110$ is given by $\sqrt{P_1}u_2 + \sqrt{P_2}v_2$, where $u_2\in \{1, -1\}$ and $v_2\in \{e^{j\pi/4}, e^{j5\pi/4}\}$. The positions of the superimposed symbols can be shown in Fig.~\ref{fig_mid_branch}. According to Fig.~\ref{fig_mid_branch}, the minimum Euclidean distance $d_{\mathrm{mid}}$ is given by
\begin{align*}
    d_{\mathrm{mid}}^2 &= \min\{\delta_4^2,\delta_5^2\} \notag\\
    &= (2\!-\!\sqrt{2})P_2 + \min\left\{0, 4P_1 + 2\sqrt{P_1P_2}\left(\sqrt{2} - 2\right)\right\}.
\end{align*}

To summarize, the minimum Euclidean distance between the diverging-and-merging paths is given by
\begin{align}\label{eq_d_dm}
    d_{\mathrm{D\&M}}^2 =& d^2_{\mathrm{diverge}} + d^2_{\mathrm{mid}} + d^2_{\mathrm{merge}}\notag\\
    =& \left(6-\sqrt{2}\right)P_2 + 8P_1 - 8\sqrt{2P_1P_2} \notag\\
    &+ \min\left\{0, 4P_1 + 2\sqrt{P_1P_2}\left(\sqrt{2} - 2\right)\right\}.
\end{align}

\subsection{Free Distance}
The free distance of $T_1 \otimes T_2$ is determined by finding the minimum of $d_\mathrm{parallel}$ and $d_\mathrm{D\&M}$, i.e.,
\begin{align}\label{eq_d_free}
    d_{\mathrm{free}}^2 =& \min\{d_{\mathrm{parallel}}^2, d_{\mathrm{D\&M}}^2\}\notag\\
    =& \min\!\left\{\!4P_1, \!4\left(\sqrt{P_2} - \sqrt{P_1}\right)^2, \left(6-\sqrt{2}\right)P_2 + 8P_1\right.\notag\\ 
    &\left.\!- 8\sqrt{2P_1\!P_2} \!+\! \min\!\left\{\!0,\! 4P_1 \!+\! 2\sqrt{P_1\!P_2}\left(\!\sqrt{2}\! -\! 2\right)\!\right\}\!\right\}.
\end{align}

The optimal powers can be derived by maximizing the free distance, i.e.,
\begin{align}\label{eq_0.24}
    \left[P_1^*, P_2^*\right] &= \arg\max_{P_1, P_2} d_{\mathrm{free}}^2,\  \mathrm{s.t.}\ P_1 + P_2 \le P,
\end{align}
where $P$ is the total transmit power. According to \eqref{eq_d_free}, one can derive that $d_{\mathrm{free}}$ is maximized when $4P_1 = \left(6-\sqrt{2}\right)P_2 + 8P_1 - 8\sqrt{2P_1P_2}$, which then results in $\frac{P_1^*}{P_2^*} = \left(\frac{2\sqrt{2} - \sqrt{2+\sqrt{2}}}{2}\right)^2 \approx 0.2404$. Besides, to combat the channel noise, $P_1 + P_2$ should be maximized. As a result, $P_1^* = \frac{0.2404}{1+0.2404} P \approx 0.1938P$ and $P_2^* \approx 0.8062P$.

While we presented the results for a two-user scenario with 8-PSK 4-state TCM, our approach can be generalized to any TCM.

\section{Simulation Results}
In this section, we present the simulation results of the 8-PSK 4-state trellis-coded NOMA (TC-NOMA), TCMA, and the uncoded NOMA (UC-NOMA) with 4-PSK. We ensure a fair comparison among these schemes since the TCM is implemented without consuming extra bandwidth compared with the uncoded modulation~\cite{ungerboeck1987trellis}. In our simulation, we employ bit error ratio (BER) as the measure of performance. In the uncoded NOMA, the maximum likelihood detection is employed. We also present the results for the TCMA where the signals for Users~1 and 2 are modulated by the identical trellis shown in Fig.~\ref{fig_4_state_trellis} but differentiated by constellation~\cite{aulin1999trellis}. In TCMA, the constellation used by one user is the other user's constellation rotated by $\pi/8$. In contrast to the trellis-coded NOMA, the transmitted signal in TCMA is given by $\sqrt{\left(P_1+P_2\right)/2}[a_1(n) + a_2(n)]$, which ensures a fair comparison by using the same sum transmit power.
\begin{figure}[t b]
	\centering
	\includegraphics[width=3.3in]{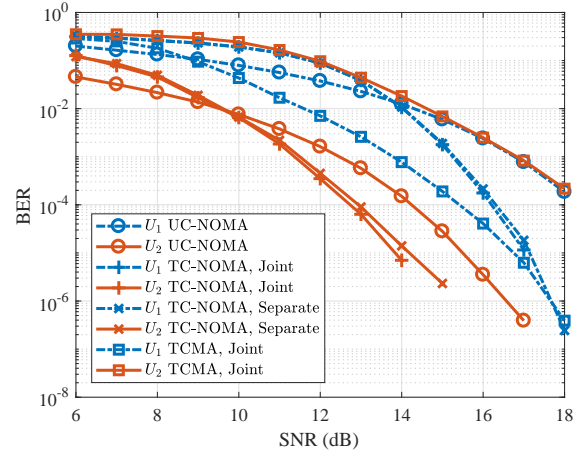}
	\caption{BER vs. SNR for TCMA, uncoded and trellis-coded NOMA when $P_1 = 0.1$, $P_2 = 1$, $|h_1|^2 = 2$, $|h_2|^2 = 1$.}
	\label{fig_P1=0.1}
\end{figure}
\begin{figure}[t b]
	\centering
	\includegraphics[width=3.3in]{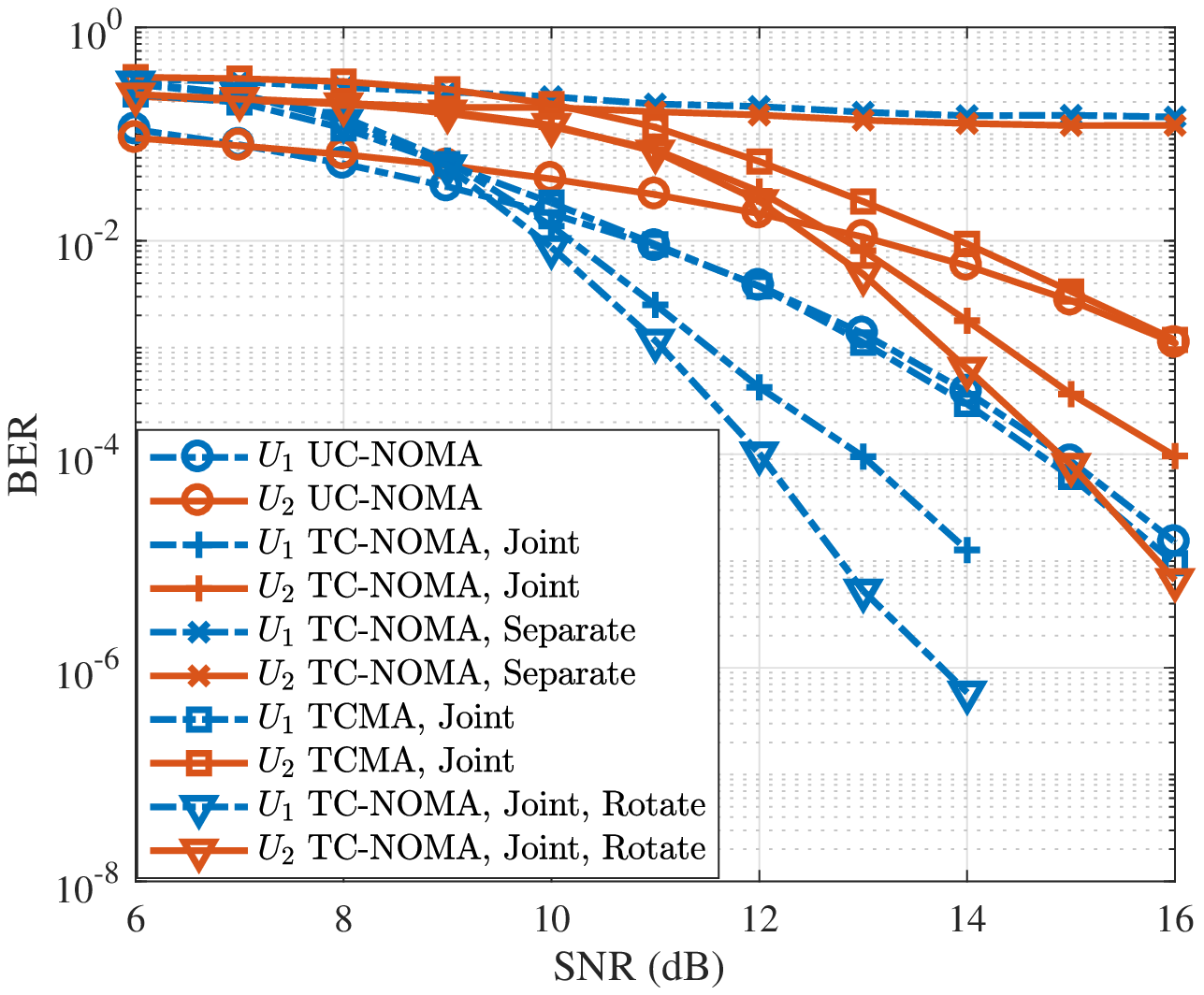}
	\caption{BER vs. SNR for TCMA, uncoded and trellis-coded NOMA when $P_1 = 0.3$, $P_2 = 1$, $|h_1|^2 = 2$, $|h_2|^2 = 1$.}
	\label{fig_P1=0.3}
\end{figure}
\begin{figure}[t b]
	\centering
	\includegraphics[width=3.3in]{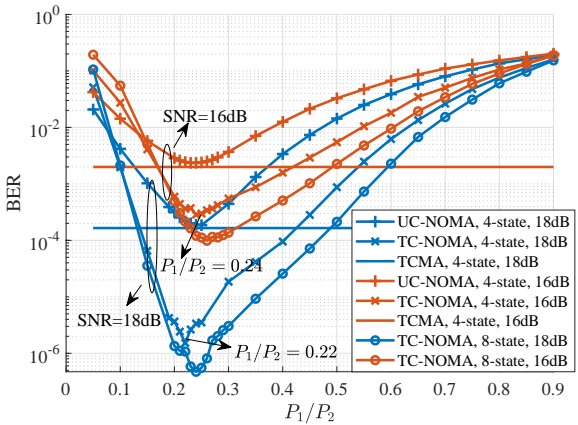}
	\caption{BER vs. $P_1/P_2$ for TCMA, uncoded and trellis-coded NOMA schemes at users employing the joint detection when $P_1 + P_2 = 1$.}
	\label{fig_power_allocation}
\end{figure}

First, we show the BER as a function of SNR for NOMA and TCMA schemes in Figs.~\ref{fig_P1=0.1} and \ref{fig_P1=0.3} when $P_2 = 1$ and $P_1 = 0.1$ or 0.3, respectively. SNR is given by $\frac{1}{\sigma^2}$ where $\sigma^2$ is the variance of noise. For $P_1 = 0.1$ or 0.3, it is manifested that at high-SNR, similar to conventional TCM~\cite{ungerboeck1987trellis}, the trellis-coded NOMA using the joint detection outperforms the uncoded NOMA. Besides, the trellis-coded NOMA using the separate detection achieves a similar performance to that using the joint detection when $P_1 = 0.1$. In contrast, there is a huge gap between the BER curves of the separate detection and those of the joint detection when $P_1 = 0.3$. This is because of the severe inter-user interference when detecting two user's signals separately and the error propagation problem in SIC. 
Furthermore, using the joint detection, the trellis-coded NOMA outperforms TCMA at high-SNR in Figs.~\ref{fig_P1=0.1} and \ref{fig_P1=0.3}. Moreover, in the trellis-coded NOMA, the signals of different users can also employ different constellations. The curves with ``TC-NOMA, Joint, Rotate'' in Fig.~\ref{fig_P1=0.3} are for the case where the constellation used by one user is the other user's constellation rotated by $\pi/8$. The trellis-coded NOMA with constellation rotation achieves a better performance compared with the trellis-coded or uncoded NOMA without constellation rotation and TCMA. It can be explained intuitively by considering how the constellation rotation affects the Euclidean distance between superimposed symbols. According to Figs.~\ref{fig_parallel_dist} and \ref{fig_diverge_merge_branch}, the minimum Euclidean distance may increase if the constellation of User~1's signal rotates by $\pi/8$, which then improves the performance. 

Fig.~\ref{fig_power_allocation} shows how the average BER changes with the power ratio $P_1/P_2$ using the joint detection at SNRs 16dB and 18dB for the 4-state trellis in Fig.~\ref{fig_4_state_trellis} and the 8-state trellis in Fig.~12.8 of \cite{benedetto1999principles}. It is shown that the minimum BER is achieved when $P_1/P_2 \approx 0.25$ for the uncoded NOMA and the 8-state trellis-coded NOMA. The optimal power ratio for the 4-state trellis-coded NOMA is 0.24 for SNR=16dB and 0.22 for SNR=18dB, which are close to the optimal power ratio of 0.2404 derived in Section~\ref{sec_power_ratio}. Moreover, the trellis-coded NOMA in its best case scenario outperforms the uncoded NOMA in its best case scenario. Besides, the performance of TCMA does not change with $P_1/P_2$. By choosing the proper powers, the trellis-coded NOMA outperforms TCMA.

\section{Conclusions}
In this letter, we study the trellis-coded NOMA and propose a joint detection method based on the tensor product of trellises. Besides, we derive the optimal power allocation between the two users by maximizing the free distance of the tensor product trellis. Simulation results demonstrate that the trellis-coded NOMA outperforms the uncoded NOMA and TCMA using an appropriate power allocation. The study of the trellis-coded NOMA systems with more than two users is our future work.

%
%
%


\ifCLASSOPTIONcaptionsoff
  \newpage
\fi



{\small
\bibliographystyle{IEEEtran}
\bibliography{IEEEabrv,IEEEexample}}

\begin{thebibliography}{10}
\providecommand{\url}[1]{#1}
\csname url@samestyle\endcsname
\providecommand{\newblock}{\relax}
\providecommand{\bibinfo}[2]{#2}
\providecommand{\BIBentrySTDinterwordspacing}{\spaceskip=0pt\relax}
\providecommand{\BIBentryALTinterwordstretchfactor}{4}
\providecommand{\BIBentryALTinterwordspacing}{\spaceskip=\fontdimen2\font plus
\BIBentryALTinterwordstretchfactor\fontdimen3\font minus
  \fontdimen4\font\relax}
\providecommand{\BIBforeignlanguage}[2]{{%
\expandafter\ifx\csname l@#1\endcsname\relax
\typeout{** WARNING: IEEEtran.bst: No hyphenation pattern has been}%
\typeout{** loaded for the language `#1'. Using the pattern for}%
\typeout{** the default language instead.}%
\else
\language=\csname l@#1\endcsname
\fi
#2}}
\providecommand{\BIBdecl}{\relax}
\BIBdecl

\bibitem{wang2018non}
Q.~Wang, R.~Zhang, L.-L. Yang, and L.~Hanzo, ``Non-orthogonal multiple access:
  A unified perspective,'' \emph{{IEEE} Wireless Commun.}, vol.~25, no.~2, pp.
  10--16, Apr. 2018.

\bibitem{liu2017downlink}
X.~Liu and H.~Jafarkhani, ``Downlink non-orthogonal multiple access with
  limited feedback,'' \emph{{IEEE} Trans. Wireless Commun.}, vol.~16, pp.
  6151--6164, Sep. 2017.

\bibitem{choi2016power}
J.~Choi, ``Power allocation for max-sum rate and max-min rate proportional
  fairness in {NOMA},'' \emph{{IEEE} Commun. Lett.}, vol.~20, no.~10, pp.
  2055--2058, Oct. 2016.

\bibitem{vanka2012superposition}
S.~Vanka, S.~Srinivasa, Z.~Gong, P.~Vizi, K.~Stamatiou, and M.~Haenggi,
  ``Superposition coding strategies: Design and experimental evaluation,''
  \emph{{IEEE} Trans. Wireless Commun.}, vol.~11, no.~7, pp. 2628--2639, Jul.
  2012.

\bibitem{zhang2011unified}
R.~Zhang and L.~Hanzo, ``A unified treatment of superposition coding aided
  communications: Theory and practice,'' \emph{{IEEE} Commun. Surveys Tuts.},
  vol.~13, no.~3, pp. 503--520, Third Quarter 2011.

\bibitem{nikopour2013sparse}
H.~Nikopour and H.~Baligh, ``Sparse code multiple access,'' in \emph{Proc.
  {IEEE} PIMRC}, London, UK, Sep. 2013, pp. 332--336.

\bibitem{aulin1999trellis}
T.~Aulin and R.~Espineira, ``Trellis coded multiple access ({TCMA}),'' in
  \emph{Proc. {IEEE} ICC}, Vancouver, BC, Canada, Jun. 1999, pp. 1177--1181.

\bibitem{di2019tcm}
B.~Di, L.~Song, Y.~Li, and G.~Y. Li, ``{TCM-NOMA}: Joint multi-user codeword
  design and detection in trellis coded modulation based {NOMA} for beyond
  {5G},'' \emph{{IEEE} J. Sel. Topics Signal Process.}, vol.~13, no.~3, pp.
  766--780, Jun. 2019.

\bibitem{jafarkhani1999design}
H.~Jafarkhani and V.~Tarokh, ``Design of successively refinable trellis-coded
  quantizers,'' \emph{{IEEE} Trans. Inf. Theory}, vol.~45, no.~5, pp.
  1490--1497, Jul. 1999.

\bibitem{jafarkhani1999multiple}
------, ``Multiple description trellis-coded quantization,'' \emph{{IEEE}
  Trans. Commun.}, vol.~47, no.~6, pp. 799 -- 803, Jun. 1999.

\bibitem{viterbi1971convolutional}
A.~Viterbi, ``Convolutional codes and their performance in communication
  systems,'' \emph{IEEE Trans. Comm. Tech.}, vol.~19, no.~5, pp. 751--772, Oct.
  1971.

\bibitem{brannstrom2002iterative}
F.~Brannstrom, T.~M. Aulin, and L.~K. Rasmussen, ``Iterative detectors for
  trellis-code multiple-access,'' \emph{{IEEE} Trans. Commun.}, vol.~50, no.~9,
  pp. 1478--1485, Sep. 2002.

\bibitem{ungerboeck1987trellis}
G.~Ungerboeck, ``Trellis-coded modulation with redundant signal sets part {I}:
  Introduction,'' \emph{{IEEE} Commun. Mag.}, vol.~25, no.~2, pp. 5--11, Feb.
  1987.

\bibitem{benedetto1999principles}
S.~Benedetto and E.~Biglieri, \emph{Principles of digital transmission: with
  wireless applications}.\hskip 1em plus 0.5em minus 0.4em\relax Springer
  Science \& Business Media, 1999.

\end{thebibliography}
\end{document}